\documentstyle[psfig,referee]{mn2e}
\input{psfig.sty}

\newif\ifAMStwofonts

\def\Mesz{M\'esz\'aros~}
\def\Pacz{Paczy\'nski~}
\def\Kluz{Klu\'zniak~}
\def\p{$e^\pm \;$}
\begin{document}

\title[Erupting fireballs]
        {Precursors and $e^\pm$ pair loading from erupting fireballs}   
\author[Ramirez-Ruiz, MacFadyen \& Lazzati]
        {Enrico Ramirez-Ruiz$^{1,2}$, Andrew I. MacFadyen$^{2}$ and
        Davide Lazzati$^{1}$
\\${\bf 1.}$ Institute of Astronomy, Madingley Road, Cambridge, CB3
        0HA, UK.
\\${\bf 2.}$ Department of Astronomy and Astrophysics, University of
      California, Santa Cruz, CA 95064, USA.}

\date{}

\maketitle

\label{firstpage}

\begin{abstract}
Recent observations suggest that long-duration $\gamma$-ray bursts and
their afterglows are produced by highly relativistic jets emitted in
core-collapse explosions.  As the jet makes its way out of the stellar
mantle, a bow shock
runs ahead and a strong thermal precursor is produced as the shock
breaks out. Such erupting fireballs 
produce a very bright $\gamma$-ray precursor as they interact with the
thermal break-out emission. The prompt $\gamma$-ray emission
propagates ahead of the fireball before it becomes optically thin,
leading to \p pair loading and radiative acceleration of the external
medium. The detection of such precursors would offer the possibility of
diagnosing not only the radius of the stellar progenitor and the
initial Lorentz factor of the collimated fireball, but also the density
of the external environment.

\end{abstract}


\section{Introduction}

A generic scheme for a cosmological $\gamma$-ray burst (GRB) model has
emerged in the last few years (see M\'esz\'aros 2001 for a
review). According to this scheme the observed $\gamma$-rays are
emitted when a relativistic energy flow is converted to
radiation (Rees \& \Mesz 
1992; 1994; \Pacz \& Xu 1994; Katz 1994; Sari \& Piran 
1995). Possible forms of the energy flow are kinetic energy of
relativistic particles or electromagnetic Poynting flux (Rees 1999). 
This energy
must be converted to radiation in an optically thin region, as the
observed bursts are not thermal. The ultimate energy source of this
relativistic outflow is the gravitational energy release associated
with temporary mass accretion onto a black hole, which results either
from the collapse of a massive rotating star (Woosley 1993; \Pacz
1998; MacFadyen \& Woosley 1999; MacFadyen, Woosley \& Heger 2001) or
from a compact merger (Lattimer \& Schramm 1976; Eichler at al. 1989;
Ruffert et al. 1997; \Kluz \& Lee 1998; Lee \& \Kluz 1999).

For the long burst afterglows localized so far, the host galaxies show
signs of ongoing star formation activity necessary for the presence of
young, massive progenitor stars (Fruchter et al. 1999). The physical
properties of the afterglows (Kulkarni et al. 1998; Fruchter 1999;
Ramirez-Ruiz, Trentham \& Blain 2002),
their locations at a few kpc from the centre of the host galaxies
(Bloom, Kulkarni \& Djorgovski 2001), the presence of iron line
features (GRB 991216, Piro et al. 2000; GRB990705, Amati et al. 2000),
and the evidence of a supernova component detected several weeks after
the burst (GRB980326, Bloom et al. 1999; GRB970228, Reichart 1999; GRB
000911, Lazzati et al. 2001), all give strong support to the idea that the
most common GRBs are linked to the cataclysmic collapse of massive
stars into black holes. In this case, the $\gamma$-rays are thought to
be produced in shocks occuring after the relativistic jet has broken
free from the stellar envelope, whose density is 
reduced along the rotation axis due to an early phase of accretion
(MacFadyen \& Woosley 1999 and MacFadyen, Woosley \& Heger 2001;
hereinafter MW99 and MWH01 respectively).   A strong
terminal wave breaking out of the envelope is expected to produce a
transient thermal emission that should appear as a precursor signal
prior to the observed GRB (Colgate 1974; Chevalier 1982; MWH01; \Mesz
\& Waxman 2001).

In this paper we explore the interaction of such erupting
fireballs with the shock break-out emission. We show that a substantial
fraction of the fireball energy can be converted into a collimated,
bright $\gamma$-ray precursor via the Compton drag process (Lazzati et
al. 2000 and Ghisellini et al. 2001; hereinafter L00 and G01
respectively). The prompt
$\gamma$-rays would propagate ahead of the fireball before it becomes
optically thin, leading to \p pair production and thus an associated
deposition of momentum into the external medium (Madau \& Thompson
2000 and Thompson \& Madau 2000; hereinafter MT00 and TM00
respectively).  
In this scenario, the effects of \p pair production can be substantial,
increasing the radiative efficiency of the blast wave and modifying
the early dynamics of the fireball (in contrast with Dermer \&
B\"ottcher 2000; Madau, Blandford \& Rees 2000; M\'esz\'aros, Ramirez-Ruiz \&
Rees 2001 and Belobodorov 2002; all of whom consider the effects of such
mechanisms beyond the radius at which the relativistic fireball becomes
optically thin to scattering). We suggest that detailed observation
of this prompt emission provides a potential tool for diagnosing the
radius of the stellar progenitor and the initial Lorentz factor of the
fireball. It also provides a means for probing the external environment
surounding the stellar progenitor and the radial distance at which the
observed $\gamma$-rays are produced. We assume $H_0 = 65\,\, {\rm km}
{\rm s}^{-1} {\rm Mpc}^{-1}$, $\Omega_{\rm m}=0.3$, $\Omega_{\Lambda}=0.7$.

\section{Erupting Fireballs}

Numerical simulations of rotating helium stars in which iron core
collapse does not produce a successful traditional neutrino-powered
explosion (MW99; Aloy et al. 2000; MWH01) have identified a range of
stellar progenitors and initial conditions in which a jet would not be
able to break free from its stellar cocoon. This is expected in a
large fraction of cases where the stellar envelope is too thick, for
example in stars with small radiative mass-loss. A highly relativistic
jet is likely to escape if 
the star loses its hydrogen envelope before collapsing and if the jet
produced by the accretion maintains its energy for longer than it
takes the jet to reach the surface of the star (MW99). Otherwise,
acceleration of the explosion debris to a sufficiently high Lorentz
factor ($>10^{2}$, M\'esz\'aros
\& Rees  1997) is unlikely and an asymmetric supernovae like SN 1998bw
may result (MWH01). 

\begin{figure*}
\psfig{figure=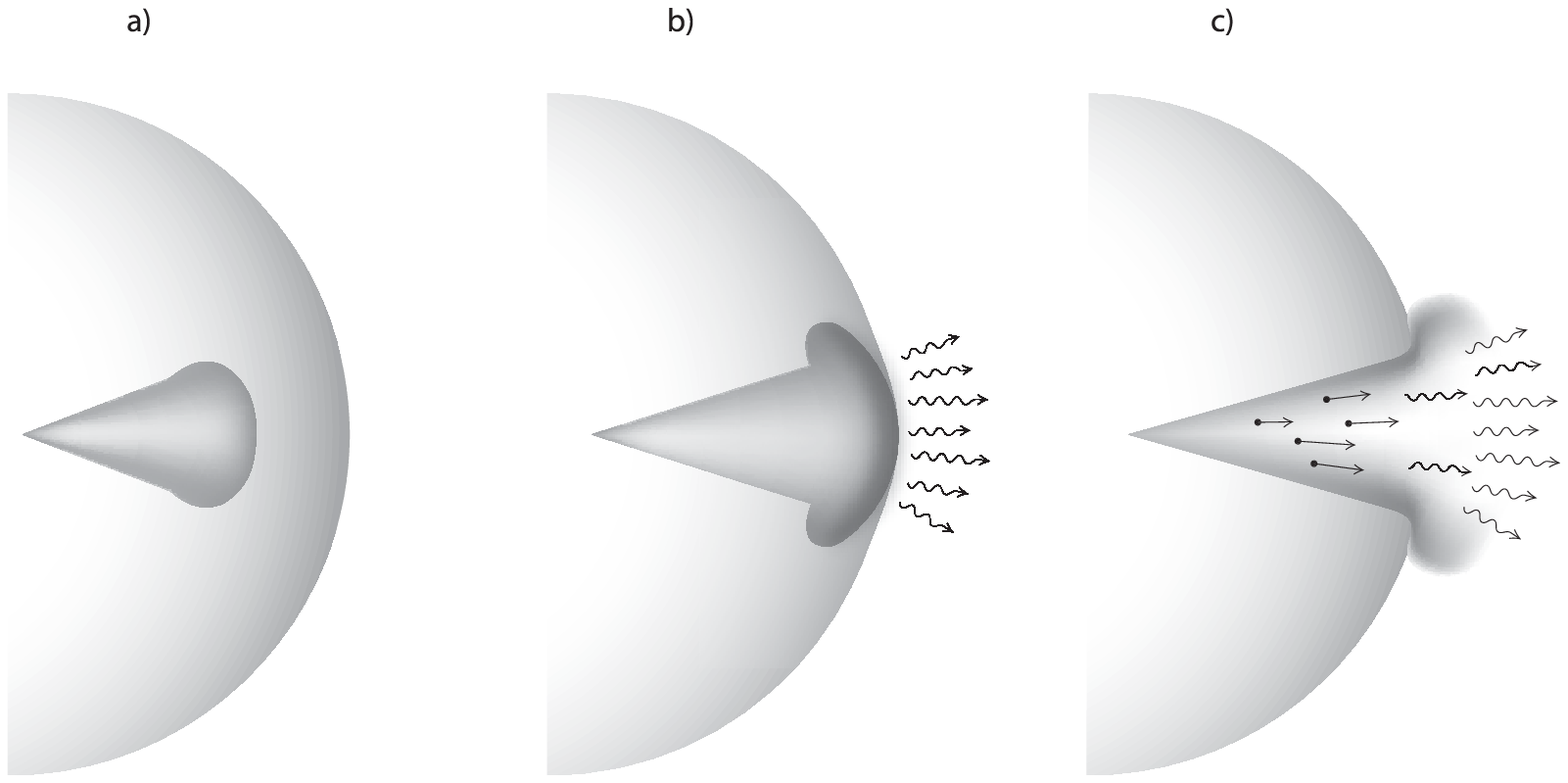,angle=0,width=0.95\textwidth}
{\caption{Diagram illustrating the propagation of the jet through the
stellar mantle. Initially, the jet is unable to move the envelope
material to a speed comparable to its own and thus is abruptly
decelerated. As the jet propagates a bow shock runs ahead of it
(a). The bow shock of the jet will both heat material and cause it to
expand sideways. A strong thermal precursor is produced as the shock
breaks through the stellar surface and exposes the hot shocked
material (b).  The fireball escapes the stellar envelope and interacts
with very dense soft photon emission (c), converting the fireball bulk
energy into radiation with a remarkably high efficiency.}
\label{fig1}}
\end{figure*}

A collimated fireball propagating inside a funnel cavity would be
stopped by the envelope when its momentum flux is insufficient to accelerate
the impacted stellar mantle to a speed comparable to its own. The jet
would then be stalled at a distance $\approx \sqrt{L_{j}/(\Omega
\rho_{\rm env} v_{j}^3)}$ (Wheeler et al. 2000), where $L_{j}$ is the
total luminosity of the jet, $\Omega$ its collimation solid angle and
$\rho_{\rm env}$ the density along the rotation axis of the star. At this
distance, the relativistic jet is abruptly decelerated to $\Gamma
\approx 1$.  In order to break through the star, the energy injected
into the envelope, $E_j=L_{j}\Delta t_{j}$, should be enough to
unbind the impacted envelope material. Thus, inside a rotating massive
star whose core has collapsed (leading to a central black hole), gas
fall-back would drive for a time $\Delta t_{j}=r_*/\bar{v_{\rm h}} \approx
10^3 r_{*, 13}\; {\rm s}$  a
slowly advancing standoff shock inside the envelope\footnote{The 
shocked jet plasma and the shocked stellar plasma propagate at
a subrelativistic velocity $\Gamma_{\rm h} \approx 1 \ll \Gamma_{j}$ up to 
the H envelope radius. During the propagation in the H envelope, the
density drops and the jet can accelerate to relativistic speed (MWH01,
\Mesz \& Waxman 2001). Here we adopt the convention $Q
= 10^x\,Q_x$, using cgs units.} (MW99; MWH01; see
Fig.~\ref{fig1}a), where the average
speed of the jet head $\bar{v_{\rm h}}$ is about c/2 (Aloy et al. 2000;
Zhang et al., 2001, in preparation).
For small opening angles $\theta \ll 1$, the thickness of
the shocked plasma shell is\footnote{Calculated by balancing the particle flux
across the standoff shock with the tangential flux of particles
leaving the cylinder of height $\Delta$.} $\Delta \approx
\sqrt{3}\theta r_{*}/8$, where $r_{*}$ is the stellar
radius (\Mesz \& Waxman 2001). The bow shock of the jet will heat the
shocked plasma, converting a large fraction of its internal energy
into radiation (Fig.~\ref{fig1}b). The high opacity of the plasma shell would
cause the radiation to thermalize with a black body emission at a
temperature\footnote{This estimate assumes that the fireball
propagates in a funnel along the rotation axis of the star, which is
supported  by rotation in the transverse direction and thus 
jet sideways expansion inside the stellar envelope does not occur.  
Alternatively, if sideways expansion of the jet material
takes place before leaving the stellar surface,  most of the energy
output could be deposited into a cocoon of
relativistic plasma surrounding the jet (Ramirez-Ruiz, Celotti \& Rees
2001, in preparation). In this case, the shocked
plasma internal energy converted to radiation is smaller and thus the
black body radiation temperature may be lower.} (\Mesz \& Waxman 2001)
\begin{equation}
T_{\rm p} \approx 0.3 \left[ {E_{j,51} \over r_{*,13}^2 \Delta t_{j,2}
\theta_{-1}^2} \right]^{1/4} {\rm keV}.  
\label{eq:t}
\end{equation}
This emission would not be appreciably beamed. 
Once the strong shock wave breaks the stellar surface, the temperature
of the shocked plasma decreases roughly as a power-law  $T(t)
\approx T_{\rm p}(1+ 3ct/\Delta)^{-1/3} $. These analytical estimates are
consistent with  numerical results (MWH01).

A relativistic collimated fireball may result if the engine operates
for a sufficiently long time to allow the standoff shock (i.e. the jet
head) to break out of the surface of the star.  High Lorentz factors
($\Gamma > 100$) can be achieved if the jet continues to be
powered after jet break out.
The column of stellar material
pushed ahead of the jet ($0.1-1 {\rm M}_\odot$) escapes the star and
expands sideways, leaving a decreasing amount of material ahead of the
jet. This allows the hot, low-density jet plasma with
moderately relativistic bulk velocity to accelerate to Lorentz factors
determined by the energy loading per baryon in the jet (see Zhang et
al. 2001, in preparation). The jet remains physically beamed due to
relativistic effects. The shocked plasma temperature at the time at
which the fireball 
crosses the surface ($t \approx \Delta/c$) is $T_{\rm s}\approx T_{\rm
p}/4^{1/3}$. Thus, the erupting fireball escapes the stellar envelope
while interacting with very dense soft photon emission with typical
energy $\Theta_{\rm s}=k T_{\rm s}/(m_e c^2)$ (Fig.~\ref{fig1}c). A fraction
$\approx {\rm min}(1,\tau_{j})$ of the photons are scattered by the
inverse Compton effect to energies $\approx 2\Gamma_0^2\Theta_{\rm
s}$, where $\tau_{j}$ is the Thomson optical depth of the collimated
fireball and we assume that a constant $\Gamma_0$ has been reached at
the stellar surface. The radius of transparency of the fireball is
\begin{equation}                                                    
r_{\tau} \approx 2 \times 10^{14}
\theta_{-1}^{-1}\Gamma_{0,2}^{-1/2}E_{j,51}^{1/2}\; {\rm cm}. 
\end{equation} 
Due to relativistic aberration, the scattered photons propagate in a
narrow $1/\Gamma$ beam. The net amount of energy $E_{\rm ic}$
extracted by the Compton drag process is $\approx  \Gamma^2{\rm
min}(1,\tau_{j}) \Delta \pi (\theta r)^2 a T_{\rm s}^4$, where $\Delta
\pi (\theta r)^2$ is the volume filled by the soft photon
radiation. The Compton drag process can be very efficient  in
extracting energy from the collimated fireball 
\begin{equation}
\xi={E_{\rm ic} \over E_{j}} \approx \theta_{-1}r_{*,13} \Delta t_{j,2}^{-1}
\Gamma_{0,2}^2\;\;\;\;\;\;\;\;\;\;\;r\;<\;r_{\tau} 
\end{equation}  
even for jets erupting from stars with
small radii $r_{*} < 10^{12}$ cm. Note that the Compton drag process
limits the maximum speed of expansion so that $\xi < 1$ (L00). 
When the fireball becomes
transparent, the amount of scattered photons is correspondingly
reduced, and the process becomes less efficient. Each seed photon is
boosted by $\approx \Gamma^2$ in frequency, yielding a spectrum peaking
at
\begin{equation}
h \nu \approx 2\Gamma_0^2(3kT_{\rm s}) \approx  11 \Gamma_{0,2}^2
{E_{j,51}^{1/4} \over r_{*,13}^{1/2} \Delta t_{j,2}^{1/4} \theta_{-1}^{1/2} }
(1+z)^{-1} {\rm MeV}. 
\end{equation}
The observed variability time scale is related to the typical size
$\Delta$ of the shocked plasma region containing the thermal photon
field, its curvature $r_*$, and the mean free path $\lambda$ of a
photon inside the fireball.  The observed time-scale is hence given by
$\max( {\lambda \over c}, {r_* \over c\Gamma_0^2},{\Delta \over
c\Gamma_0^2})$, where
$\lambda\sim10^{10}\,(r_{*,13}\theta_{-1})^2\,\Gamma_{0,2}\,t_{j,2}\,E_{j,51}^{-1}$~cm.
For large Lorentz factors, the duration is dominated by the mean free
path term, while for slow fireballs the curvature is  more important.
Taking into account the conservation of the number of photons and the
increase by a factor $2\Gamma_0^2$ of the photon energy, the peak
luminosity of the boosted component is
\begin{equation}
L_{\rm ic} \approx 2^{-5/3}\Gamma_0^2\,{{\delta t_{\rm b}}\over{\delta t_{\rm
ic}}}\, {{4\pi}\over{\Omega}}\,L_{\rm b}
\end{equation}
where subscripts ${\rm ic}$ and ${\rm b}$ refer to the boosted and
break out
quantities, respectively. When the boosted duration $\delta t_{\rm ic}$ is
dominated by curvature effects, we have $L_{\rm
ic}\approx 0.3\theta\Gamma_0^4\,L_{\rm b}$.
The above observational signatures would be present even if $\Gamma_0$
is low, as expected for stars in which the jet either fails to
maintain sufficient collimation or loses a significant amount of
energy before breaking out of the star.
 
\section{Observable Precursor Effects}
The prompt thermal signal emerging from shock break-out would precede
by $\approx \Delta/c \approx 6\;\theta_{-1} r_{*,13}$ seconds the
prompt emission produced through the Compton drag process. In most
cases, and especially if we consider the BATSE [20-600]~keV spectral
window, the break out emission is too soft to be detected (see
equation ~\ref{eq:t}). For compact star progenitors ($r_*<10^{12}$~cm),
however, the break out emission could be detectable with instruments like
Ginga and the {\it Beppo}SAX wide field cameras (see Section 3.1).
The boosted  component, on the other hand, could be  hard  
and may be difficult to disentangle from the internal shock emission. In fact,
this up-scatter emission should appear as a transient signal   
$(r_{\gamma}-r_{*})/(2c\Gamma_0^2)$ seconds prior to the main burst, where
$r_\gamma=\max({r_\tau},r_{\rm int})$ is the radius at which the
$\gamma$--rays are produced and $r_{\rm int}$ 
is the radius of internal shocks.  As shown by equation 2, the radius of
transparency of the fireball is likely to happen at some distance from
the stellar surface and thus the time delay between the up scatter
emission and the burst is given by $r_{\tau}/(2c\Gamma_0^2) \approx
0.3\;\theta_{-1}^{-1} E_{j,51}^{1/2} \Gamma_{0,2}^{-5/2} {\rm s}$. If
some bursts are characterised by low-intermediate Lorentz factors,
precursors with typical photon energies 
$h\nu\sim100\,\Gamma_{0,1}^2$~keV should have been observed by BATSE
$\sim 100\,\Gamma_{0,1}^{-5/2}$~s before the main event. Such
precursors may have indeed been observed (Koshut et al. 1995).

\begin{figure*}
\psfig{figure=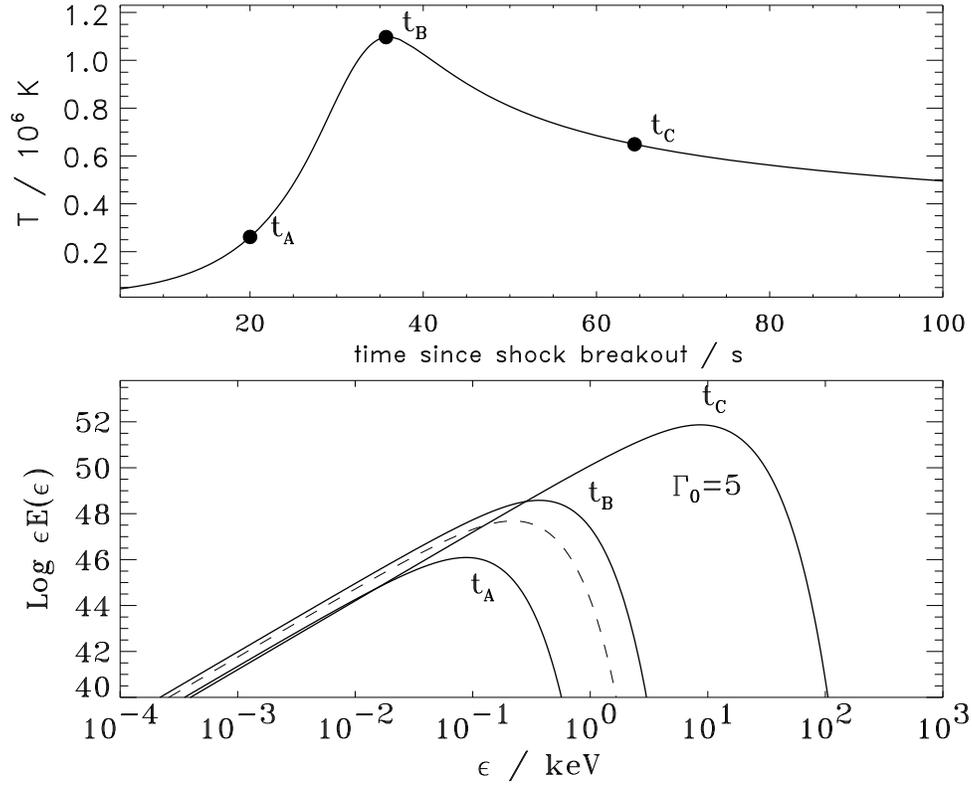,width=0.95\textwidth}
{\caption{Shock break-out and Compton-dragged prompt emission in core
collapse explosions. The stellar progenitor A01 of MWH01 is used as
the initial model, a pre-supernova star of initial mass $M_{\rm i}=25\;{\rm
M}_\odot$ that ended its life with a low-density hydrogen envelope of
6.57 ${\rm M}_\odot$ and a stellar radius of $\approx 9 \times
10^{13}$ cm.  Upper panel: the effective temperature as a function of
time since shock break out. Lower panel: examples of the the
black-body spectra during the initial phase ($t_{\rm A}$ and $t_{\rm
B}$) before the impact of the jet ($E_{4\pi} \approx 10^{54}$ and
$\theta$ = 0.6). The fireball interacts with the dense thermal emission
at a time $t_{\rm c}= \Delta/c$, boosting photons by $\approx
2\Gamma_0 \approx 2(5)^2$ in frequency. The dashed line corresponds to
the emission at $t_{\rm c}$ neglecting the Compton drag process.}
\label{fig2}}
\end{figure*}

\begin{figure*}
\psfig{file=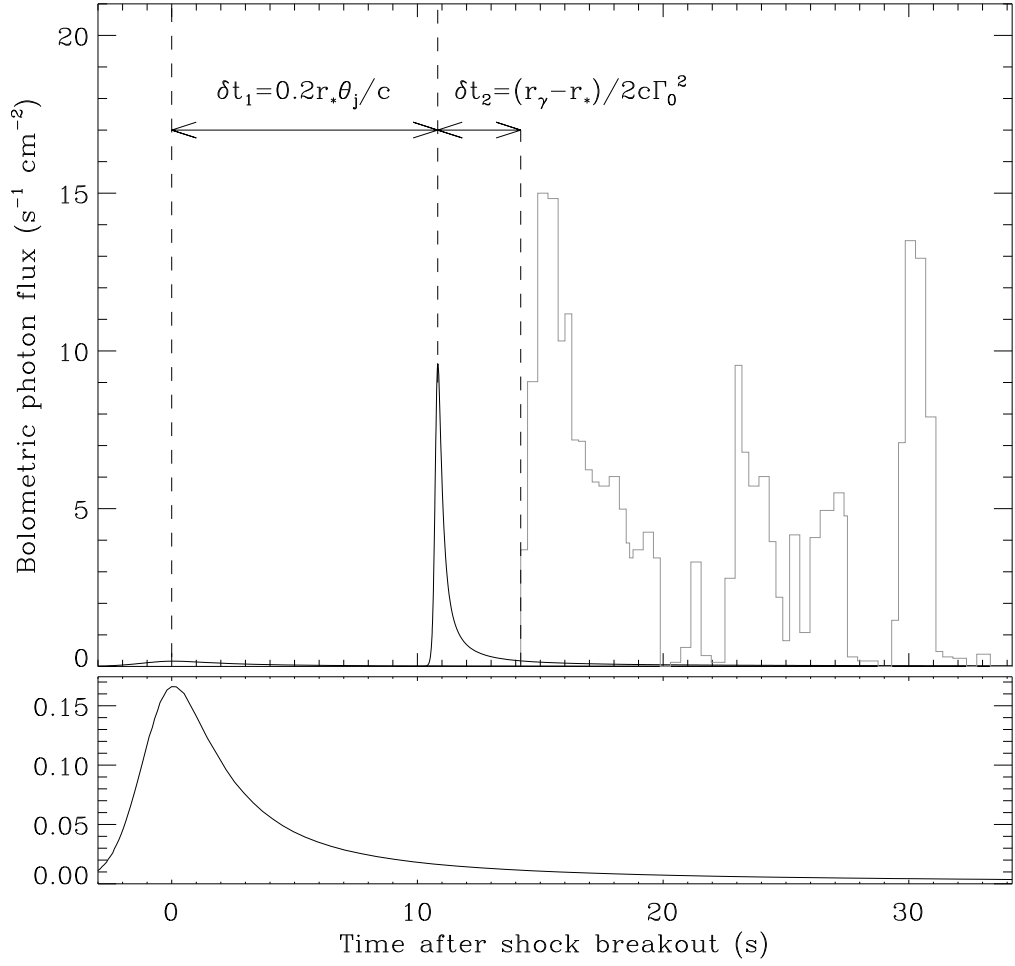,width=0.9\textwidth}
\caption{{Lightcurve of the break-out, Compton-drag and internal shock 
components of a GRB. The lightcurve is computed to match the
observation of GRB900126: $r_*=8\times10^{11}$~cm, $\theta=0.9$,
$E_j=3\times10^{51}$~erg, $\Delta t_j=20$~s, $\Gamma_0=8$ and for a
redshift $z=1$. The main panel shows all the three components of the
lightcurve while the lower panel shows a close up of the break out component.}
\label{fig:lcu}}
\end{figure*}

Fig.~\ref{fig2} shows the break out emission of a fireball
propagating through the stellar envelope expected at the end of the
evolution of a 25 ${\rm M}_\odot$ main-sequence star (model A01 of
MWH01).  The effect of a strong, spherically symmetric shock that
breaks through the stellar surface is calculated using the
non-relativistic hydrodynamics KEPLER code. The equivalent isotropic
energy of the fireball pushing the stellar envelope is\footnote{While
these calculations are one 
dimensional, they should roughly simulate the conditions experienced
by stellar regions within solid angles that are being pushed by these
equivalent isotropic energies, as described by MWH01.}  $E_{4\pi}
\approx 10^{54}$ erg. Due to the low
efficiency of semi-convective mixing, the main sequence star ends its
life with a relatively large hydrogen envelope ($r_{*}
\approx 9 \times 10^{13}$ cm). In this case, the short duration of the jet, 
which is limited by a gas fall-back time onto the central black hole,
prevents the fireball from acquiring a large bulk Lorentz factor
($\Gamma_0 \approx 5$). An asymmetric supernovae and a weak GRB 
may result from such explosions.  Fig.~\ref{fig:lcu} shows the three
component bolometric lightcurve for a mildly beamed burst with
intermediate Lorentz factor. The break-out, boosted and internal shock
components are shown (see the caption for more details).

\subsection{GRB 900126}

On 1990 January 26, the Ginga experiment discovered X-ray emission in
the 2--10 keV energy range $\sim 10$ s before the onset of a
$\gamma$-ray event (Murakami et al. 1991).  The $\gamma$-ray signal
shows two distinct peaks, separated by $\sim 6$ s, both of which
have rise times of $\approx 1.5$ s. The peak energy values in the
$\gamma$-ray emission vary from $\approx 120$~keV in the first peak to
$\approx 80$~keV in the second. The spectrum of the precursor X-ray
emission can be described by that of a black body with a temperature
$kT=1.58 \pm 0.26$~keV and a flux
$F\sim2.5\times10^{-9}$~erg~cm$^{-2}$~s$^{-1}$ (Murakami et
al. 1991). The peak luminosity of the first spike is
$\sim9.5\times10^{-6}$~erg~cm$^{-2}$~s$^{-1}$, i.e. $\sim2600$ times
brighter than the thermal precursor.

Knowing the flux and observed temperature of the thermal precursor, it
is possible to estimate the radius $r_{\rm s}$ of the emitting surface as a
function of redshift.  In the redshift range $0.5 <z <10$, we obtain
$10^{11} < r_{\rm s} <3\times10^{11}$~cm. In the framework of the
shock break out, the radius of the emitting surface is given by
$r_{\rm s} \approx r_{*}\theta$ and the thermal  precursor should
precede the boosted emission by $0.2r_*\theta(1+z)/c\approx2(1+z)$~s.
The first peak of the
$\gamma$--ray emission is indeed observed several seconds after the
thermal precursor. If this second peak is interpreted as the Compton
boosted emission described above, we can derive the Lorentz factor
from the peak frequency of the second pulse and the temperature of the
precursor, yielding $\Gamma_0 \approx 8$. The expected ratio of
luminosity between the thermal precursor and the boosted pulse would
then be $\approx 2500\theta$, to be compared with the measured value of
$\approx 2600$. This would consequently imply that the fireball of
GRB900126 was only moderately beamed and the radius of the progenitor
star was $r_*\approx 10^{11}$~cm. The expected shock break out temperature
is (see equation~\ref{eq:t}) $\approx 1 (E_{j,51}/\Delta
t_{j,2})^{1/4}$~keV, so that a comparison with the observed $T\approx
1.6(1+z)$~keV implies a 
total energy of few $\approx 10^{52}$~erg. The fluence of the burst was
${\cal F}=4\times10^{-5}$~erg~cm$^{-2}$ which, for a $z=1$ burst with
a mild ($\theta\approx 1$) beaming, corresponds to
$E_j=5\times10^{52}$ (Fig.~\ref{fig:lcu}). On the contrary, if the
softening between the first and second $\gamma$-ray peaks is caused by
the same mechanism that is responsible for both of these emissions
(i.e. internal shocks), the lack of detection of the Compton drag
transient in the 1.5--400 keV band would imply that $\Gamma_0 \ge 20$.

\section{Fireball Kinematics and \p pair loading}

Here we quantitatively estimate the predicted spectrum of the prompt
transient that arises from the interaction of a fireball with the
break-out emission using the framework established by G01 and give a
simplified discussion of the generic effects of pair formation.  We
assume that the photon field inside the stellar cocoon is not very
dense and thus the Compton drag process is not efficient until the
fireball reaches the shocked plasma region\footnote{The Compton drag
effect when the fireball propagates in a funnel embedded in a very
dense photon bath has been invoked by L00 and G01. In this scenario,
the fireball experiences efficient Compton drag as it reaches the end
of the funnel. At this end, the soft photon energy is determined by
the stellar surface temperature.}.

Beyond the stellar surface and in the region where the fireball
remains optically thick ($r<r_{\tau}$), the total energy emitted by
the fireball through the Compton drag process over a distance $dr$ is
\begin{equation}
dE(r)={\pi \over 2}\theta^2r^2a
T_{s}^4\left({r \over r_{*}}\right)^{-\alpha}\Gamma^2(r)dr, 
\end{equation}
where the dynamics of the
fireball obey 
\begin{equation}
M_{j}c^2 {d\Gamma(r) \over dr} = {\pi \over 2}\theta^2r^2 a
T_{s}^4 \left( {r \over r_{*}}\right)^{-\alpha}\Gamma^2(r)
\end{equation}
as long as most of the fireball
energy is lost in the Thomson regime (G01). For simplicity, as the
scattering rate is $\propto (1-\beta \cos \theta) \propto
(r/r_{*})^{-2}$ and the radiation energy $U(r) \propto
(r/r_{*})^{-2}$, we assume\footnote{It is also
plausible that $U(r) \times$ (scattering rate) can have a more complex
profile for which $2\;< \alpha <\; 4$, as described in G01} $\alpha \approx 4$.

The resulting spectrum and the dynamics of the fireball can be
strongly affected by the production of \p pairs through
$\gamma$-$\gamma$ interactions. \p pairs can be produced by Compton
drag photons interacting with the isotropic break-out emission or
with each other. This latter interaction can be between photons
among the beam or with a seed fraction of back-scattered
photons. $\gamma$-$\gamma$ collisions among the beam occur between
photons of equal age (cascade shower; Burns \& Lovelace 1982; Svensson
1986) and can only affect the high energy tail of the spectrum if
$\Gamma_0 \Theta_{s} > {1 \over 3}$ (Svensson 1987, G01).

\begin{figure*}
\psfig{figure=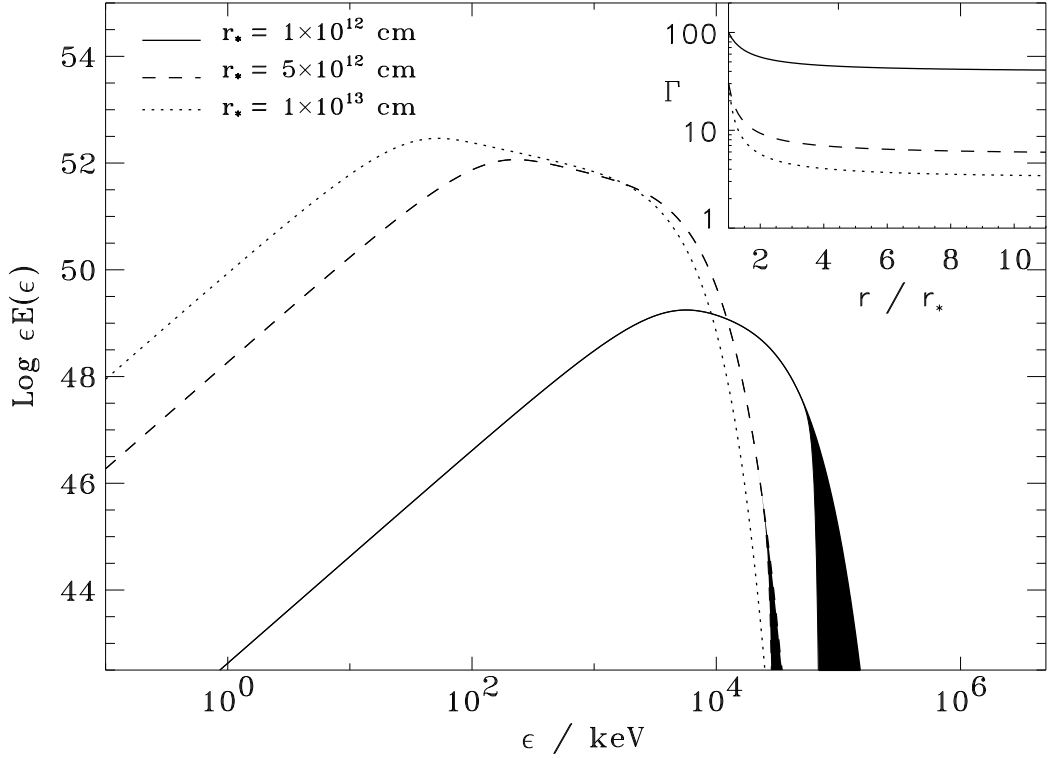,width=0.95\textwidth}
{\caption{Examples of spectra produced by the interaction of a
collimated fireball with the break-out emission. The inset panel shows
the corresponding bulk Lorentz factor. The model parameters are:
$E_{j}= 10^{51}$ erg (solid line), $5 \times 10^{52}$ erg (dashed and
dotted lines); $\theta$ = 0.2 (solid line) and 0.1 (dashed and dotted
lines). In all cases we assume $\Delta t_{j} \approx 10^{2}$ s and
$\alpha$ = 4. The shaded regions correspond to the emission absorbed by
$\gamma-\gamma$ interaction between beam photons and break-out
radiation.}
\label{fig3}}
\end{figure*}

The very high energy emission produced by the Compton drag can also
interact with the soft break out photons and produce \p
pairs. However, the number of target photons able to interact with the
high energy $\gamma$-rays to produce pairs strongly decreases beyond
$r_{*}$. To illustrate the importance of this interaction, we
calculate the observed Compton-drag spectrum, taking into account
this photon-photon absorption. The observed spectrum
($r_*\;<\;r\;<\;r_{\tau}$) is then given by
\begin{equation}
E(\epsilon)={\pi^2 \over 4}m_e c^2 \left({m_e c \over h} \right)^3
\int_{r_{*}}^{r_{\tau}} {
r^2(r/r_*)^{-\alpha} e^{-\tau_{\gamma \gamma}} \epsilon^3 \over
\Gamma^6(r) (e^{\epsilon / \Theta_{c} }-  1 )} dr,   
\end{equation} 
where $\Theta_{c}=2\Gamma^2\Theta_{s}$, $\epsilon$ is the photon
energy in units of $m_e c^2$ and $\tau_{\gamma \gamma}$ is the
photon-photon optical depth of Compton drag photons interacting with
the soft break-out emission (see equations 11, 12 and 13 of G01). In
Fig.~\ref{fig3} we show three examples of both the predicted spectrum and the
$\Gamma$ profiles corresponding to different values of the stellar
envelope and the total fireball energy. The resulting spectrum is
calculated with and without the absorption term $e^{-\tau_{\gamma
\gamma}}$ (shaded region). If the Compton drag process is efficient
($\xi \propto r_{*}$, see equation 3), the fireball decelerates and
the observed spectrum is the convolution of all the locally emitted
spectra. Beyond $r_{*}$ the photon density is strongly reduced, thus
further decreasing both the efficiency of the Compton drag process and
the amount of energy absorbed in $\gamma$-$\gamma$ collisions.

As can be seen in Fig.~\ref{fig3}, the radiation absorbed by $\gamma$-$\gamma$
interaction between beam photons and break-out radiation is
small. However, the resulting spectra are hard, with a significant
fraction of the energy above the $\gamma\gamma \rightarrow$ \p
formation energy threshold; and a high compactness parameter can
result in additional pairs being formed outside the stellar radius
(MT00; TM00).

When the integrated flux of photons of energy $\epsilon > 1$ is large,
each side-scattered photon deposits its entire momentum, along with
the momentum of the interacting beam photon, into an \p pair. The
pairs (together with the ions) start being accelerated to
$\Gamma_{\mu} \ge 1$ (provided that the pairs remain coupled to the
baryons; TM00).  Particles are accelerated away from the stellar
surface as long as $\Gamma_{\mu} < \Gamma_{\rm eq}$, where 
\begin{equation}
\Gamma_{\rm eq}(r) \approx 3^{1/4}{r \over r_{*}}
\end{equation} 
is the equilibrium Lorentz factor close to
the extended source (MT00).  Bulk motion starting with
$\Gamma_{\mu}(r) >\Gamma_{\rm eq} $ is quickly decelerated to
$\Gamma_{\mu} \approx \Gamma_{\rm eq}$. This saturation of the Lorentz
factor is due to the drag caused by the aberration into the forward
hemisphere of blue-shifted photons as seen in the particle
rest-frame. Large values of $\Gamma_{\mu}$ can only be obtained for
particles injected at distances beyond the critical radius 
\begin{equation}
r_c \approx 0.8(lm_e/ \mu)^{1/4}r_{*},
\end{equation}
at which drag caused by the radiation field is negligible ($l m_e/
\mu = L \sigma_T/ (4 \pi \mu c^3 r_{*})$ is the rescaled compactness
parameter and $\mu$ is the mean mass per scattering charge, see
MT00). 
At a distance $r >r_c$ in front of the radiation source, \p + ions can
be accelerated to a maximum value of $\Gamma_{\mu}$  satisfying
\begin{equation}
\Gamma_{\mu}\approx \left\{ 
\begin{array}{ll}
 \left( {3 l m_e r_{*} \over 4 r \mu} \right)^{1/3} & 
 \;\;\;\;\;\;{2 \over 3}\Gamma_{\mu}^2c\delta t_{\rm ic} > r \\
 {l \delta t_{\rm ic} c r_{*} \over r^2 }   & 
 \;\;\;\;\;\; {\rm otherwise}. \\ 
\end{array} 
\right. 
\end{equation}

\begin{figure*}
\psfig{figure=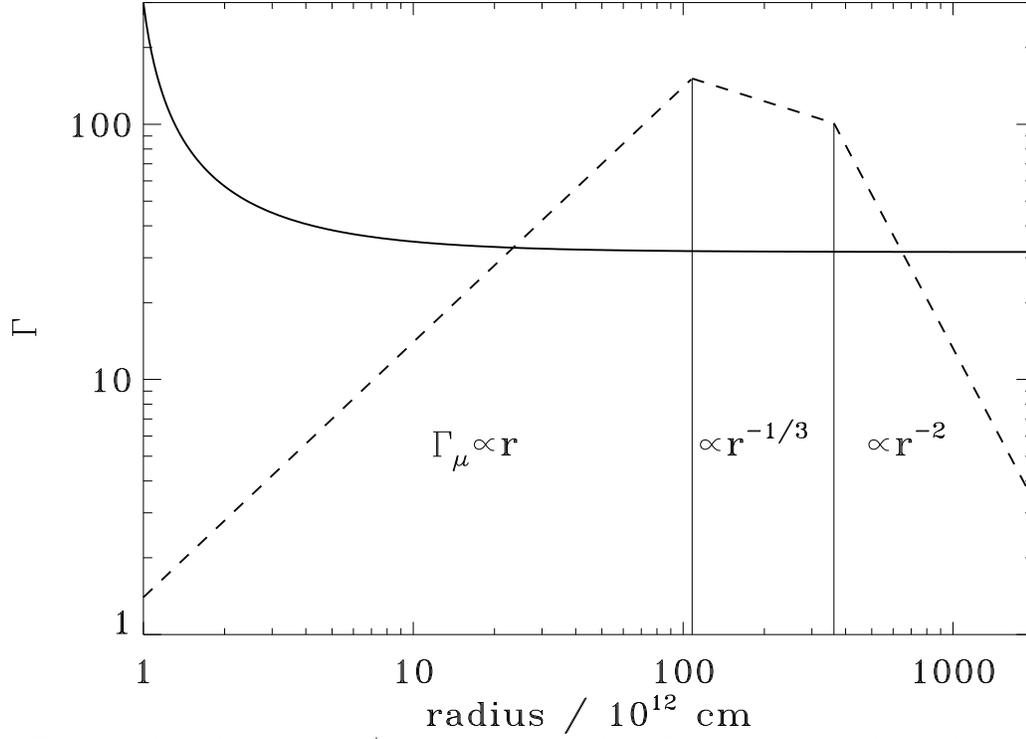,width=0.9\textwidth}
{\caption{Plot of the maximum (\p + ions) bulk Lorentz factor
$\Gamma_{\mu}$ as a function of radius (dashed line). The model
parameters are: $E_{j}= 10^{51}$ erg, $\theta$= 0.1, $r_{*}=10^{12}$
cm, $\Delta t_{j} \approx 10^{2}$ s, $\delta t_{\rm ic}=\lambda / c$
and $\mu \approx m_p$. The solid 
line gives the profile of the bulk Lorentz factor of the fireball
decelerated by the Compton drag mechanism. When pairs are produced in
sufficient numbers, the mean mass per scattering charge drops to $\mu
\approx m_e$ and acceleration can be  more efficient.} 
\label{fig4}}
\end{figure*}

The maximum Lorentz factor $\Gamma_{\mu}$ is shown in Fig.~\ref{fig4}.
The \p pair loading process depends only on the ``seed'' $\gamma$-ray
photon, while its manifestations are consequences of the external
density and on the initial bulk Lorentz factor (\Mesz et al. 2001). 
The external baryon
density $n_{\rm ext}$ determines the optical depth that can be built
up through back-scattering and pair multiplication. For a fixed
Lorentz factor $\Gamma_0$, the external density determines when the
outer shock and the reverse shock become important and whether this
happens within the radius already polluted with pairs (MT00; TM00;
\Mesz et al. 2001).

For small values of $\xi$, the \p pair-rich external medium would
carry less energy and inertia than the erupting fireball itself, and
therefore  start to decelerate at a smaller radius than the
collimated fireball, so that the latter would overtake it (\Mesz et
al. 2001). On the other hand, if the Compton drag process dominates,
the pair energy can exceed that of the collimated fireball (after
kinetic energy has been extracted from it), substantially altering the
usual properties of the deceleration shocks and the afterglow
emission. In this case, the afterglow would be dominated by the
emission of the accelerated medium and not by the decelerated
fireball.  The lepton/proton fraction in the ejecta can then be much
larger than normal, causing the radiation to be much softer than in the
usual model, because the same energy density has to be shared among a
larger number of particles. Pair production can also increase the
optical depth outside of the shocks, and both the Compton drag and the
internal shock emission may be modified by Comptonization. We plan to
investigate these possibilities and their consequences for the
predicted prompt and afterglow emissions in future work.

\section{Summary}
Many massive stars produce supernovae when forming neutron stars in
spherically symmetric explosions, but some may fail neutrino energy
deposition, forming a black hole in the centre of the star and
possibly a GRB (MW99). One expects various outcomes ranging from GRBs
with large energies and durations, to asymmetric, energetic supernovae
with weak GRBs. The prompt transients produced by the Compton drag of
the shock break-out emission would provide a natural test to
distinguish between these different stellar explosions. The detection
of these prompt multi-wavelength signatures would be a test of the
collapsar model; and the precise measurement of the time delay between
emissions would constrain the dimensions of the stellar progenitor,
the Lorentz factor of the fireball and the radius of the burst
emitting region ($r_{\rm int}$ or $r_{\tau}$).

This very hard prompt emission would propagate ahead of the collimated
fireball loading the external medium with \p pairs. In most early
discussions (TM00; Madau et al. 2000; \Mesz et al 2001; Belobodorov
2002), the concern has been raised that \p pair loading in low-density
environments is rather inefficient, converting only a few percent of
the bulk motion energy into \p pairs. We have shown here that the Compton drag
mechanism can be an effective catalyst for converting bulk motion
energy into $\gamma$-rays close to the stellar surface. Numerous \p pairs
can then be produced as some of the photons in the beam are
backscattered and interact with other incoming photons. The process
discussed here suggests that the \p pairs can play a substantial role
in both the dynamics of the fireball and the nature of the afterglow
emission, as they are produced well before the fireball becomes
optically thin. This suggests that if GRBs are the outcome of the
collapse of massive stars involving a relativistic fireball jet,
the time structure, dynamics and efficiency of the prompt and
afterglow emissions may have a more complex dynamic than the standard
models suggest.

\section*{Acknowledgements}
We thank A. Celotti, P. Madau, P. M\'esz\'aros, G. Morris, 
S. Woosley, A. Heger and W. Zhang for many helpful conversations. We are
particularly grateful to M. J. Rees for numerous insightful comments and
suggestions. ERR thanks the UCSC Department of Astronomy and Astrophysics
for its hospitality and acknowledges support from CONACYT, SEP and the
ORS foundation. A.M. acknowledges support from DOE ASCI (B347885).

\bsp

\label{lastpage}

\newpage

\newpage

\newpage

\newpage

\newpage

\end{document}